\documentclass{PoS}

\usepackage[varg]{txfonts} 
\usepackage[latin1]{inputenc}

\RequirePackage{longtable}
\RequirePackage{tabularx}
\RequirePackage{dcolumn}

\newcommand{\mev}{\mbox{MeV}}
\newcommand{\gev}{\mbox{GeV}}
\newcommand{\be}{\begin{equation}}
\newcommand{\ee}{\end{equation}}
\newcommand{\bea}{\begin{eqnarray}}
\newcommand{\eea}{\end{eqnarray}}
\newcommand{\bdm}{\begin{displaymath}}
\newcommand{\edm}{\end{displaymath}}

\title{Hadronic light-by-light scattering in the muon $g-2$: impact of
  proposed measurements of the $\pi^0 \to \gamma\gamma$ decay width
  and the $\gamma^*\gamma \to \pi^0$ transition form factor with the
  KLOE-2 experiment}

\ShortTitle{Hadronic light-by-light scattering in the muon $g-2$:
  impact of KLOE-2}

\author{\speaker{Andreas Nyffeler}\\
        Regional Centre for Accelerator-based Particle Physics\\
        Harish-Chandra Research Institute\\
        Chhatnag Road, Jhusi, Allahabad - 211019, India\\ 
        E-mail: \email{nyffeler@hri.res.in}}

\abstract{ 
The calculation of the hadronic light-by-light scattering contribution
to the muon $g-2$ currently relies entirely on models. Measurements of
the form factors which describe the interactions of hadrons with
photons can help to constrain the models and reduce the uncertainty in
$a_\mu^{{\rm had.\ LbyL}}= (116 \pm 40) \times 10^{-11}$.  In the
numerically dominant pion-exchange contribution, the form factor
${\cal F}_{{\pi^0}^\ast\gamma^\ast\gamma^\ast}((q_1 + q_2)^2, q_1^2,
q_2^2)$ with an off-shell pion enters. In general, measurements of the
transition form factor ${\cal F}_{\pi^0\gamma^\ast\gamma}(Q^2) \equiv
{\cal F}_{{\pi^0}^\ast\gamma^\ast\gamma^\ast}(m_{\pi}^2, -Q^2, 0)$ are
only sensitive to a subset of the model parameters. Thus, having a
good description for ${\cal F}_{\pi^0\gamma^\ast\gamma}(Q^2)$ is only
necessary, not sufficient, to determine $a_\mu^{{\rm LbyL}; \pi^0}$.

Simulations have shown that planned measurements at KLOE-2 should be
able, within one year of data taking, to determine the
$\pi^0\to\gamma\gamma$ decay width to 1\% statistical precision and
the $\gamma^\ast\gamma\to\pi^0$ transition form factor ${\cal
  F}_{\pi^0\gamma^\ast\gamma}(Q^2)$ for small space-like momenta,
$0.01~\mbox{GeV}^2 \leq Q^2 \leq 0.1~\mbox{GeV}^2$, to 6\% statistical
precision in each bin. Note that in the two-loop integral for the
pion-exchange contribution the relevant regions of momenta are in the
range $0-1.5~\gev$.

With the decay width $\Gamma_{\pi^0\to\gamma\gamma}^{\rm PDG}$
$[\Gamma_{\pi^0\to\gamma\gamma}^{\rm PrimEx}]$ and current data for
the transition form factor ${\cal F}_{\pi^0\gamma^\ast\gamma}(Q^2)$,
the error on $a_\mu^{{\rm LbyL}; \pi^0}$ is $\pm 4 \times 10^{-11}$
[$\pm 2 \times 10^{-11}$], not taking into account the uncertainty
related to the off-shellness of the pion. Including the simulated
KLOE-2 data reduces the error to $\pm (0.7 - 1.1) \times
10^{-11}$. For models like VMD, which have only few parameters that
are completely determined by measurements of ${\cal
  F}_{\pi^0\gamma^\ast\gamma}(Q^2)$, this represents the total
error. But maybe such models are too simplistic. In other models,
e.g.\ those based on large-$N_c$ QCD, parameters describing the
off-shell pion dominate the uncertainty in $a_{\mu; {\rm
    large-}N_c}^{{\rm LbyL}; \pi^0} = (72 \pm 12) \times 10^{-11}$.  }

\FullConference{The 7th International Workshop on Chiral Dynamics\\
                August 6 - 10, 2012\\
                Jefferson Lab, Newport News, Virginia, USA}

\begin{document}

\section{Introduction}
\label{intro}

The anomalous magnetic moment of the muon $a_\mu$ provides an
important test of the Standard Model (SM) and is potentially sensitive
to contributions from New Physics. For some time now a deviation is
observed between the experimental measurement and the SM prediction,
$a_\mu^{\rm exp} - a_\mu^{\rm SM} \sim (250-300) \times 10^{-11}$,
corresponding to $3 - 3.5$~standard
deviations~\cite{JN_09,recent_estimates}. Hadronic effects dominate
the uncertainty in the SM prediction of $a_\mu$. In contrast to the
hadronic vacuum polarization in the $g-2$, which can be related to
data, the estimates for the hadronic light-by-light (LbyL) scattering
contribution $a_\mu^{{\rm had.\ LbyL}} = (105 \pm 26) \times
10^{-11}$~\cite{PdeRV_09} and $a_\mu^{{\rm had.\ LbyL}} = (116 \pm 40)
\times 10^{-11}$~\cite{Nyffeler_09,JN_09} rely entirely on
calculations using hadronic models which employ form factors for the
interaction of hadrons with photons. Some
papers~\cite{Schwinger_Dyson_GdeR_12} yield a larger central value and
a larger error of $(150 \pm 50) \times 10^{-11}$, see also further
analyses and partial evaluations of hadronic LbyL scattering in
Refs.~\cite{ZA+B_ZA,LbyL_recent_papers}.  To fully profit from future
planned $g-2$ experiments with a precision of $15 \times 10^{-11}$,
these large model uncertainties have to be reduced. Maybe lattice QCD
will at some point give a reliable number, see
Ref.~\cite{LbyL_Lattice_2012}. Meanwhile, experimental measurements
and theoretical constraints of the relevant form factors can help to
constrain the models and to reduce the uncertainties in $a_\mu^{{\rm
    had.\ LbyL}}$.

In most model calculations, pion-exchange gives the numerically
dominant contribution.  The relevant momentum regions\footnote{For
  attempts to visualize the relevant momentum regions in hadronic LbyL
  scattering for the pseudoscalars and for other contributions, see
  Refs.~\cite{KN_PRD_2002,BP_2007,ZA+B_ZA}.}  for all the light
pseudoscalars, $\pi^0, \eta, \eta^\prime$, can be inferred from
Table~\ref{tab:cutoffdependence}, where we list, for different models
of the form factor, the results obtained for a given UV cutoff
$\Lambda$ in the 3-dimensional integral representation derived in
Ref.~\cite{JN_09}. The cutoff bounds the length of the two Euclidean
momenta, $|Q_i| < \Lambda, i=1,2$ in the two-loop integral. The third
integration variable is the angle between the two 4-vectors $Q_i$. The
off-shell LMD+V model~\cite{KN_EPJC_2001,Nyffeler_09} is based on
large-$N_c$ QCD matched to short-distance constraints from the
operator product expansion. For the vector-meson dominance (VMD)
model, the vector meson mass has been obtained by fitting
data~\cite{TFF_data} for the pseudoscalar-photon transition form
factors. The model parameters are the same as in Ref.~\cite{Nyffeler_09}.

\begin{table}[b] 
\begin{center} 
{\small 
\begin{tabular}{|c|c|c|c|c|c|}
\hline 
$\Lambda$ & \multicolumn{3}{c|}{$\pi^0$} & $\eta$ & $\eta^\prime$
\\{} 
[GeV]     & \multicolumn{1}{c}{LMD+V $(h_3 = 0)$} &
\multicolumn{1}{c}{LMD+V $(h_4 = 0)$} & \multicolumn{1}{c|}{VMD}  
          & VMD & VMD \\
\hline 
0.25 & 14.8~(20.6\%) & 14.8~(20.3\%) & 14.4~(25.2\%) 
     & 1.76~(12.1\%) & 0.99~(7.9\%)  \\ 
0.5  & 38.6~(53.8\%) & 38.8~(53.2\%) & 36.6~(64.2\%) 
     & 6.90~(47.5\%) & 4.52~(36.1\%) \\
0.75 & 51.9~(72.2\%) & 52.2~(71.7\%) & 47.7~(83.8\%) 
     & 10.7~(73.4\%) & 7.83~(62.5\%) \\
1.0  & 58.7~(81.7\%) & 59.2~(81.4\%) & 52.6~(92.3\%)
     & 12.6~(86.6\%) & 9.90~(79.1\%) \\
1.5  & 64.9~(90.2\%) & 65.6~(90.1\%) & 55.8~(97.8\%)
     & 14.0~(96.1\%) & 11.7~(93.2\%) \\
2.0  & 67.5~(93.9\%) & 68.3~(93.8\%) & 56.5~(99.2\%)
     & 14.3~(98.6\%) & 12.2~(97.4\%) \\
5.0  & 71.0~(98.8\%) & 71.9~(98.8\%) & 56.9~(99.9\%) 
     & 14.5~(99.9\%) & 12.5~(99.9\%) \\
20.0 & 71.9~(100\%)  & 72.8~(100\%)  & 57.0~(100\%)
     & 14.5~(100\%)  & 12.5~(100\%)  \\ 
\hline 
\end{tabular}
}
\caption{The pseudoscalar exchange contribution,
  $a_\mu^{\mathrm{LbyL;PS}}\times 10^{11}, \mbox{PS}=\pi^0, \eta,
  \eta^\prime$, for different models with a cutoff for the two
  Euclidean momenta, $|Q_i| < \Lambda, i=1,2$ in the two-loop
  integral. In brackets, the relative contribution of the total
  obtained with $\Lambda = 20~\gev$.}
\label{tab:cutoffdependence}
\end{center} 
\end{table}

For the pion the bulk of the contribution comes from the region below
$\Lambda = 1~\gev$, about 82\% for the LMD+V form factor and about
92\% for the VMD form factor. For the VMD form factor, the small
contribution from the region with momenta higher than $1~\gev$ can be
understood from the weight-functions in the integrals derived in
Ref.~\cite{KN_PRD_2002}, which peak around $0.5~\gev$, and the strong
fall-off of the VMD form factor at large momenta. For the off-shell
LMD+V form factor, the region with larger momenta is more important as
the form factor drops off less quickly and there is no damping at the
external vertex, see Ref.~\cite{Nyffeler_09}.  For the $\eta$ and
$\eta^\prime$, the peaks of the relevant weight functions in the
integrals are shifted to higher values of $|Q_i|$ and the saturation
effect only sets in around $\Lambda = 1.5~\gev$, with about 95\% of
the contribution to the total.

\begin{figure}[h]
\centerline{\includegraphics[width=.7\textwidth]{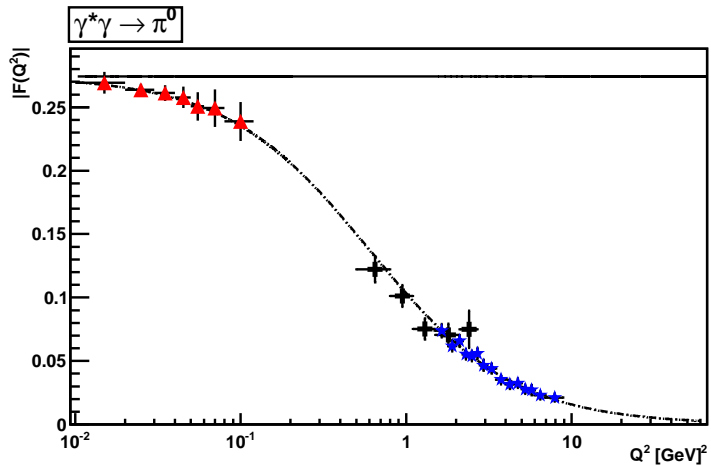}}
\caption{Simulation of KLOE-2 measurement of $F(Q^2)$ (red triangles)
  with statistical errors for $5$~fb$^{-1}$, corresponding to one year
  of data taking. The dashed line is the $F(Q^2)$ form factor
  according to the LMD+V model, the solid line is $F(0) = 1/(4\pi^2
  F_\pi)$ given by the Wess-Zumino-Witten term. Data~\cite{TFF_data}
  from CELLO (black crosses) and CLEO (blue stars) at high $Q^2$ are
  also shown for illustration.}
\label{simulation_FF_data}
\end{figure}

In Ref.~\cite{KLOE-2_impact} it was shown that planned measurements at
KLOE-2 could determine the $\pi^0\to\gamma\gamma$ decay width to 1\%
statistical precision and the $\gamma^\ast\gamma\to\pi^0$ transition
form factor ${\cal F}_{\pi^0\gamma^\ast\gamma}(Q^2)$ for small
space-like momenta, $0.01~\mbox{GeV}^2 \leq Q^2 \leq
0.1~\mbox{GeV}^2$, to 6\% statistical precision in each bin. The
simulations have been performed with the Monte-Carlo program
EKHARA~\cite{EKHARA} for the process $e^+ e^- \to e^+ e^- \gamma^*
\gamma^* \to e^+ e^- \pi^0$, followed by the decay $\pi^0 \to
\gamma\gamma$ and combined with a detailed detector simulation. The
results of the simulations are shown in
Figure~\ref{simulation_FF_data}. The KLOE-2 measurements will allow to
almost directly measure the slope of the form factor at the origin and
check the consistency of models which have been used to extrapolate
the data from larger values of $Q^2$ down to the origin.

\section{Impact of KLOE-2 measurements on $a_\mu^{\mathrm{LbyL};\pi^0}$}
\label{sec:KLOE}

Any experimental information on the neutral pion lifetime and the
transition form factor is important in order to constrain the models
used for calculating the pion-exchange contribution.  However, having
a good description, e.g.\ for the transition form factor, is only
necessary, not sufficient, in order to uniquely determine
$a_\mu^{\mathrm{LbyL};\pi^0}$. As stressed in
Ref.~\cite{Jegerlehner_off-shell}, what enters in the calculation of
$a_\mu^{{\rm LbyL}; \pi^0}$ is the fully off-shell form factor ${\cal
  F}_{{\pi^0}^*\gamma^*\gamma^*}((q_1 + q_2)^2, q_1^2, q_2^2)$ (vertex
function), where also the pion is off-shell with 4-momentum $(q_1 +
q_2)$. Such a (model dependent) form factor can for instance be
defined via the QCD Green's function $\langle VVP \rangle$, see
Ref.~\cite{Nyffeler_09} for details.  The form factor with on-shell
pions is then given by ${\cal F}_{\pi^0\gamma^*\gamma^*}(q_1^2, q_2^2)
\equiv {\cal F}_{{\pi^0}^*\gamma^*\gamma^*}(m_\pi^2, q_1^2, q_2^2)$.
Measurements of the transition form factor ${\cal
  F}_{\pi^0\gamma^\ast\gamma}(Q^2) \equiv {\cal
  F}_{{\pi^0}^\ast\gamma^\ast\gamma^\ast}(m_{\pi}^2, -Q^2, 0)$ are in
general only sensitive to a subset of the model parameters and do not
allow to reconstruct the full off-shell form factor.

For different models, the effects of the off-shell pion can vary a
lot. In Ref.~\cite{Nyffeler_09} the off-shell LMD+V form factor was
proposed and the estimate $a_{\mu; {\rm LMD+V}}^{{\rm LbyL}; \pi^0} =
(72 \pm 12) \times 10^{-11}$ was obtained. The error estimate comes
from the variation of all model parameters, where the uncertainty of
the parameters related to the off-shellness of the pion completely
dominates the total error. In contrast to the off-shell LMD+V model,
many other models, e.g.\ the VMD model or constituent quark models, do
not have these additional sources of uncertainty related to the
off-shellness of the pion. These models often have only very few
parameters, which can all be fixed by measurements of the transition
form factor or from other observables. Therefore, for such models, the
precision of the KLOE-2 measurement can dominate the total accuracy of
$a_\mu^{\mathrm{LbyL};\pi^0}$.

Essentially all evaluations of the pion-exchange contribution use for
the normalization of the form factor ${\cal
  F}_{{\pi^0}^*\gamma^*\gamma^*}(m_\pi^2, 0, 0) = 1 / (4 \pi^2
F_\pi)$, as derived from the Wess-Zumino-Witten (WZW) term. Then the
value $F_\pi = 92.4~\mev$ is used without any error attached to it,
i.e. a value close to $F_\pi = (92.2 \pm 0.14)~\mbox{MeV}$, obtained
from $\pi^+ \to \mu^+ \nu_\mu(\gamma)$~\cite{Nakamura:2010zzi}. If one
uses the decay width $\Gamma_{\pi^0 \to \gamma\gamma}$ for the
normalization of the form factor, an additional source of uncertainty
enters, which has not been taken into account in most
evaluations~\cite{Nyffeler:2009uw}. We account for this by using in
the fits:

\vspace*{-0.2cm}
\begin{itemize}
\item $\Gamma^{{\rm PDG}}_{\pi^0 \to \gamma\gamma} = 7.74 \pm
0.48$~eV from the PDG 2010~\cite{Nakamura:2010zzi}
\vspace*{-0.2cm}
\item $\Gamma^{{\rm PrimEx}}_{\pi^0 \to \gamma\gamma} = 7.82 \pm
0.22$~eV from the PrimEx experiment~\cite{Larin:2010kq}
\vspace*{-0.2cm}
\item $\Gamma^{{\rm KLOE-2}}_{\pi^0 \to \gamma\gamma} = 7.73 \pm
0.08$~eV for the KLOE-2 simulation (assuming a $1\%$ precision). 
\end{itemize}

\vspace*{-0.2cm} The assumption that the KLOE-2 measurement will be
consistent with the LMD+V and VMD models, allowed us in
Ref.~\cite{KLOE-2_impact} to use the simulations as new ``data'' and
evaluate the impact on the precision of the $a_\mu^{{\rm LbyL};
  \pi^0}$ calculation. We fit the models to the data
sets~\cite{TFF_data} from CELLO, CLEO and BaBar for the transition
form factor and the values for the decay width given above:

\vspace*{-0.2cm}
\bdm 
\label{eq:fitdatasets}
 \begin{array}{llll}
   A0: & \mbox{CELLO, CLEO, PDG} &    
   \hspace*{0.5cm}B0: & \mbox{CELLO, CLEO, BaBar, PDG} \\
   A1: & \mbox{CELLO, CLEO, PrimEx} & 
   \hspace*{0.5cm}B1: & \mbox{CELLO, CLEO, BaBar, PrimEx} \\
   A2: & \mbox{CELLO, CLEO, PrimEx, KLOE-2} & 
   \hspace*{0.5cm}B2: & \mbox{CELLO, CLEO, BaBar, PrimEx, KLOE-2}\\   
 \end{array}
\edm

\vspace*{-0.2cm} The BaBar measurement does not show the $1/Q^2$
behavior as expected from theoretical
considerations~\cite{Brodsky-Lepage} and as seen in the data of CELLO,
CLEO and Belle.  The VMD model always shows a $1/Q^2$ fall-off and
therefore is not compatible with the BaBar data. The LMD+V model has
another parameter, $h_1$, which determines the behavior of the
transition form factor for large $Q^2$. To get the $1/Q^2$ behavior,
one needs to set $h_1 = 0$. However, one can simply leave $h_1$ as a
free parameter and fit it to the BaBar
data~\cite{Nyffeler:2009uw}. Since VMD and LMD+V with $h_1 = 0$ are
not compatible with the BaBar data, the corresponding fits are very
bad and we will not include these results in the current paper. We
use two ways to calculate $a_\mu^{{\rm LbyL}; \pi^0}$: the
Jegerlehner-Nyffeler (JN) approach~\cite{Nyffeler_09,JN_09} with the
off-shell pion form factor and the Melnikov-Vainshtein (MV)
approach~\cite{Melnikov:2003xd} with the on-shell pion form factor at
the internal vertex and a constant (WZW) form factor at the external
vertex.

Table~\ref{tab:amu} shows the impact of the PrimEx and the future
KLOE-2 measurements on the model parameters and on the $a_\mu^{{\rm
    LbyL}; \pi^0}$ uncertainty. The other parameters of the (on-shell
and off-shell) LMD+V model have been chosen as in the
papers~\cite{Nyffeler_09,JN_09,Melnikov:2003xd}.  We stress that our
estimate of the $a_\mu^{{\rm LbyL}; \pi^0}$ uncertainty is given only
by the propagation of the errors of the fitted parameters in
Table~\ref{tab:amu}. We can clearly see from Table~\ref{tab:amu} that
for each given model and each approach (JN or MV), there is a trend of
reduction in the error for $a_\mu^{{\rm LbyL}; \pi^0}$ by about half
when going from A0 (PDG) to A1 (including PrimEx) and by about another
half when going from A1 to A2 (including KLOE-2):

\vspace*{-0.2cm}
\begin{itemize}
\item 
Sets A0, B0: $\delta a_\mu^{{\rm LbyL}; \pi^0} \approx 4 \times
10^{-11}$ (with $\Gamma^{{\rm PDG}}_{\pi^0 \to \gamma\gamma}$)
\vspace*{-0.3cm}
\item 
Sets A1, B1: $\delta a_\mu^{{\rm LbyL}; \pi^0} \approx2 \times
10^{-11}$
(with $\Gamma^{{\rm PrimEx}}_{\pi^0 \to \gamma\gamma}$) 
\vspace*{-0.3cm}
\item   
Sets A2, B2: $\delta a_\mu^{{\rm LbyL}; \pi^0} \approx (0.7 - 1.1)
\times 10^{-11}$ (with simulated KLOE-2 data)
\end{itemize}

\begin{table*}
 {\scriptsize 
 \begin{tabularx}{\textwidth}{|c|l|c|lll|l|}
\hline 
 Model&Data& $\chi^2/d.o.f.$ &  & Parameters && $a_\mu^{{\rm LbyL};
   \pi^0} \times 10^{11}$\\ 
 \hline
 VMD  & A0 & $6.6/19$
     & $M_V = 0.778(18)$~GeV & \hspace*{-0.2cm}$F_\pi =
 0.0924(28)$~GeV  & & $(57.2 \pm 4.0)_{JN}$\\ 
 VMD  & A1 & $6.6/19$
     & $M_V = 0.776(13)$~GeV & \hspace*{-0.2cm}$F_\pi =
 0.0919(13)$~GeV  & & $(57.7 \pm 2.1)_{JN}$\\ 
 VMD  & A2 & $7.5/27$
     & $M_V = 0.778(11)$~GeV & \hspace*{-0.2cm}$F_\pi = 0.0923(4)$~GeV
 & & $(57.3 \pm 1.1)_{JN}$\\ 
 \hline
 LMD+V, $h_1 = 0$  & A0 & $6.5/19$ 
     &  $\bar{h}_5 = 6.99(32)$~GeV$^4$ & \hspace*{-0.2cm}$\bar{h}_7 =
 -14.81(45)$~GeV$^6$ & & $(72.3 \pm 3.5)_{JN}^*$\\ 
 &  &   &                        &                            &&
 $(79.8 \pm 4.2)_{MV}$\\ 
 LMD+V, $h_1 = 0$  & A1 & $6.6/19$ 
     &  $\bar{h}_5 = 6.96(29)$~GeV$^4$ & \hspace*{-0.2cm}$\bar{h}_7 =
 -14.90(21)$~GeV$^6$ & & $(73.0 \pm 1.7)_{JN}^*$\\ 
 &  &   &                        &                            &&
 $(80.5 \pm 2.0)_{MV}$\\ 
 LMD+V, $h_1 = 0$  & A2 & $7.5/27$
     &  $\bar{h}_5 = 6.99(28)$~GeV$^4$ & \hspace*{-0.2cm}$\bar{h}_7 =
 -14.83(7)$~GeV$^6$ & & $(72.5 \pm 0.8)_{JN}^*$\\ 
 &  &   &                        &                           && $(80.0
 \pm 0.8)_{MV}$\\ 
 \hline
 LMD+V, $h_1 \neq 0$  & A0 & $6.5/18$ 
     & $\bar{h}_5 = 6.90(71)$~GeV$^4$ & \hspace*{-0.2cm}$\bar{h}_7 =
 -14.83(46)$~GeV$^6$&  
        \hspace*{-0.2cm}$h_1 = -0.03(18)$~GeV$^2$ & $(72.4 \pm 3.8)_{JN}^*$\\
 LMD+V, $h_1 \neq 0$  & A1 & $6.5/18$ 
     &  $\bar{h}_5 = 6.85(67)$~GeV$^4$ & \hspace*{-0.2cm}$\bar{h}_7 =
 -14.91(21)$~GeV$^6$&  
        \hspace*{-0.2cm}$h_1 = -0.03(17)$~GeV$^2$ & $(72.9 \pm 2.1)_{JN}^*$\\
 LMD+V, $h_1 \neq 0$  & A2 & $7.5/26$ 
     &  $\bar{h}_5 = 6.90(64)$~GeV$^4$ & \hspace*{-0.2cm}$\bar{h}_7 =
 -14.84(7)$~GeV$^6$ & 
        \hspace*{-0.2cm}$h_1 = -0.02(17)$~GeV$^2$ & $(72.4 \pm 1.5)_{JN}^*$\\
 LMD+V, $h_1 \neq 0$  & B0 & $18/35$ 
     &  $\bar{h}_5 = 6.46(24)$~GeV$^4$ & \hspace*{-0.2cm}$\bar{h}_7 =
 -14.86(44)$~GeV$^6$ & 
        \hspace*{-0.2cm}$h_1 = -0.17(2)$~GeV$^2$ & $(71.9 \pm 3.4)_{JN}^*$\\
 LMD+V, $h_1 \neq 0$  & B1 & $18/35$ 
     &  $\bar{h}_5 = 6.44(22)$~GeV$^4$ & \hspace*{-0.2cm}$\bar{h}_7 =
 -14.92(21)$~GeV$^6$ & 
        \hspace*{-0.2cm}$h_1 = -0.17(2)$~GeV$^2$ & $(72.4 \pm 1.6)_{JN}^*$\\
 LMD+V, $h_1 \neq 0$  & B2 & $19/43$ 
     &  $\bar{h}_5 = 6.47(21)$~GeV$^4$ & \hspace*{-0.2cm}$\bar{h}_7 =
 -14.84(7)$~GeV$^6$ & 
        \hspace*{-0.2cm}$h_1 = -0.17(2)$~GeV$^2$ & $(71.8 \pm 0.7)_{JN}^*$ \\
\hline
 \end{tabularx}
}
 \caption{KLOE-2 impact on the accuracy of $a_\mu^{{\rm
       LbyL}; \pi^0}$ in case of one year of data taking
   ($5$~fb$^{-1}$). The values marked with asterisk (*)
   do not contain additional uncertainties coming from
   the ``off-shellness'' of the pion.} 
 \label{tab:amu}
\end{table*}

\vspace*{-0.2cm}
This is mainly due to the improvement in the normalization of the form
factor, related to the decay width $\pi^0 \to \gamma\gamma$,
controlled by the parameters $F_\pi$ or $\bar{h}_7$, respectively, but
more data also better constrain the other model parameters $M_V$ or
$\bar{h}_5$. This trend is also visible in the last part of the Table
(LMD+V, $h_1 \neq 0$), when we fit the sets B0, B1 and B2 which
include the BaBar data.

Note that both VMD and LMD+V with $h_1 = 0$ can fit the data sets A0,
A1 and A2 for the transition form factor very well with essentially
the same $\chi^2$ per degree of freedom for a given data set.
Nevertheless, the results for the pion-exchange contribution differ by
about $20\%$ in these two models. For VMD the result is $a_\mu^{{\rm
    LbyL}; \pi^0} \sim 57.5 \times 10^{-11}$ and for LMD+V with $h_1 =
0$ it is $72.5 \times 10^{-11}$ with the JN approach and $80 \times
10^{-11}$ with the MV approach. This is due to the different behavior,
in these two models, of the fully off-shell form factor ${\cal
  F}_{{\pi^0}^*\gamma^*\gamma^*}((q_1 + q_2)^2, q_1^2, q_2^2)$ on all
momentum variables, which enters for the pion-exchange
contribution. The VMD model is known to have a wrong high-energy
behavior with too strong damping. For the VMD model, measurements of
the neutral pion decay width and the transition form factor completely
determine the model parameters $F_\pi$ and $M_V$ and the error given
in Table~\ref{tab:amu} is the total model error. Note that a smaller
error does not necessarily imply that the VMD model is better,
i.e.\ closer to reality. Maybe it is too simplistic.

We conclude that the KLOE-2 data with a total integrated luminosity of
$5$~fb$^{-1}$ will give a reasonable improvement in the part of the
$a_\mu^{{\rm LbyL}; \pi^0}$ error associated with the parameters
accessible via the $\pi^0 \to \gamma\gamma$ decay width and the
$\gamma^\ast\gamma \to \pi^0$ transition form factor. Depending on the
modelling of the off-shellness of the pion, there might be other,
potentially larger sources of uncertainty which cannot be improved by
the KLOE-2 measurements.


\section*{Acknowledgements}

\vspace*{-0.25cm}
I would like to thank my coauthors on Ref.~\cite{KLOE-2_impact} for
the pleasant collaboration on that work. I am grateful to the
organizers of Chiral Dynamics 2012 for providing a stimulating
atmosphere during the meeting and for financial support. The
participation at the workshop was made possible by funding from the
Department of Atomic Energy, Government of India, for the Regional
Centre for Accelerator-based Particle Physics (RECAPP), Harish-Chandra
Research Institute and the Heinrich-Greinacher-Stiftung, Physics
Institute, University of Bern.  


\end{document}